\documentstyle[12pt,aasms4]{article}
\newcommand{\gsim}{\mbox{\raisebox{-1.0ex}{$~\stackrel{\textstyle >}
{\textstyle \sim}~$ }}}

\newcommand{\vect}[1]{\mbox{\boldmath${#1}$}}
\newcommand{\vects}[1]{\mbox{\scriptsize\boldmath${#1}$}}
\newcommand{\lmk}{\left(}
\newcommand{\rmk}{\right)}
\newcommand{\lnk}{\left\{ }
\newcommand{\rnk}{\right\} }
\newcommand{\lkk}{\left[}
\newcommand{\rkk}{\right]}
\newcommand{\llangle}{\left\langle}
\newcommand{\rrangle}{\right\rangle}

\newcommand{\vex}{{\vect x}}

\newcommand{\vek}{{\vect k}}
\newcommand{\vev}{{\vect v}}
\newcommand{\veq}{{\vect q}}
\newcommand{\ver}{{\vect r}}
\newcommand{\vep}{{\vect p}}
\newcommand{\vesx}{{\vects x}}

\newcommand{\vesk}{{\vects k}}

\newcommand{\vesq}{{\vects q}}

\newcommand{\vexa}{{\vect x}_A}
\newcommand{\vexb}{{\vect x}_B}
\newcommand{\veqa}{{\vect q}_A}
\newcommand{\veqb}{{\vect q}_B}
\newcommand{\veva}{{\vect v}_A}
\newcommand{\vevb}{{\vect v}_B}
\newcommand{\vevab}{{\vect v}_{AB}}
\newcommand{\verab}{{\vect r}_{AB}}
\newcommand{\vepa}{{\vect p}_A}
\newcommand{\vepb}{{\vect p}_B}
\newcommand{\vevp}{{\vect v}_{\scriptsize\|}}
\newcommand{\vevv}{{\vect v}_{\perp}}
\newcommand{\rab}{r_{AB}}

\newcommand{\vvi}{v_{\perp i}}
\newcommand{\vp}{{v}_{\scriptsize\parallel}}
\newcommand{\vpi}{{v}_{\scriptsize\parallel i}}
\newcommand{\vabi}{{v}_{ABi}}
\newcommand{\deltak}{\delta_{\vects k}}
\newcommand{\xii}{\xi_i}

\newcommand{\beq}{\begin{equation}}
\newcommand{\eeq}{\end{equation}}
\newcommand{\beqa}{\begin{eqnarray}}
\newcommand{\eeqa}{\end{eqnarray}}
\newcommand{\dotD}{\dot{D}}
\newcommand{\etal}{et al.\ }
\newcommand{\Mpc}{\rm Mpc}
\newcommand{\lab}{\label}
\begin{document}

\title{EVOLUTION OF THE PAIRWISE PECULIAR VELOCITY DISTRIBUTION FUNCTION IN
LAGRANGIAN PERTURBATION THEORY\\}
\author{\sc Naoki Seto }
\affil{Department of Physics, Faculty of Science, Kyoto University,
Kyoto 606-01, Japan}
\author{\sc Jun'ichi Yokoyama}
\affil{Yukawa Institute for Theoretical Physics, Kyoto University,
Kyoto 606-01, Japan}

\abstract{The statistical distribution of the radial pairwise peculiar velocity
of galaxies is known to have an exponential form as implied by
observations and explicitly shown in $N-$body simulations.
Here we calculate its statistical distribution function using
the Zel'dovich approximation assuming that the primordial density
fluctuations are Gaussian distributed.  We show that the exponential
distribution is realized as a transient phenomena on megaparsec scales 
in the standard cold-dark-matter model.
\\
{\bf Subject headings:}
cosmology: theory --- galaxies: clustering --- gravitation
--- large-scale structure of the Universe 
--- methods: statistics and analytical  }

\date{\today}
\newpage
\baselineskip 8mm
\section{Introduction}
The pairwise velocity distribution function of galaxies plays an 
important role in translating observed clustering properties in 
redshift space to those in real space.  Since Peebles (1976)
proposed to adopt a distribution function of an exponential form, a
number of observational analyses (Davis \& Peebles 1983, Hale-Sutton
\etal 1989, Fisher \etal 1994, Marzke \etal 1995) have confirmed 
it provides a better fit than other distributions.  The results of
$N$-body simulations (Efstathiou \etal 1988, Fisher \etal 1994) 
have shown more directly
that the distribution has indeed an exponential tail with a flatter
peak.  More recently, Zurek \etal (1994, hereafter ZQSW) have shown
that the pairwise 
halo velocity distribution function has an skewed exponential shape 
with a sharp peak in their simulations of cold-dark-matter (CDM)
models with a higher resolution.

In spite of the simplicity of the functional shape, its origin had
been a mystery for a long time.  It was expected that the exponential
shape arose as a result of highly nonlinear gravitational effects,
because linear evolution preserves the initial shape of the velocity
distribution which is usually assumed to be Gaussian.

Recently Sheth (1996) and independently Diaferio and Geller (1996)
proposed a model which explains the exponential shape of the pairwise
velocity distribution function.  They consider a universe composed by
isothermal clumps whose mass function is given by Press-Schechter
(1974) theory or observation.  Assuming that velocity distribution of
galaxies in each clump is Maxwellian whose dispersion is determined by 
its mass and that each pair of galaxies of interest is contained in
a common clump, they have calculated the weighted sum of the pairwise
velocity distribution function and shown that it has indeed an
exponential tail.  Diaferio and Geller (1996) have further discussed
that if they adopt a velocity distribution function of galaxies in a
clump which is more highly peaked than Gaussian, they can reproduce
the sharp exponential core of the final pairwise velocity
distribution.  

Although their idea is interesting and attractive, 
its applicability is limited
to a highly nonlinear regime with a small separation up to $R \lesssim
1h^{-1}$ Mpc (Diaferio \& Geller 1996), 
because their approach is valid only when each pair
belongs to the same nearly virialized clump whose typical size is
around $1h^{-1}$ Mpc.  On the other hand, the results of $N-$body
simulations clearly show that the exponential
shape is kept for much larger separations up to $\sim 5.5h^{-1}$ Mpc (ZQSW)
and for scales with a small value of the two-point correlation
function $\xi \sim 0.1$ (Efstathiou \etal 1988, Fisher \etal 1994).
This motivates us to consider a different approach to explain the
exponential shape on these scales; that is, not only a highly
nonlinear effect but also a semi-nonlinear dynamics may play a role
in realizing the exponential feature.

In the present paper we calculate the pairwise velocity distribution
function in the semi-nonlinear
regime using a Lagrangian perturbation theory known
as the Zel'dovich approximation (Zel'dovich 1970), 
assuming that the primordial
density fluctuation is random Gaussian.  In this approximation each
mass element moves in a remarkably simple manner.  In particular,
peculiar velocity field remains proportional to the ``shift'' vector
which is determined by the profile of  initial density fluctuation. 
Hence, starting from Gaussian fluctuation, the statistical
distribution of one-point peculiar velocity field remains Gaussian
(Kofman \etal 1994).  
The pairwise velocity distribution of an arbitrary fixed separation, 
however, is obtained by an
appropriate weighted sum of initial distribution function and can result
in markedly non-Gaussian distributions in semi-nonlinear regime.
As a result, an exponential shape is observed in the course of
evolution as seen below.

The rest of the paper is organized as follows.  In \S 2 we formulate
the pairwise peculiar velocity distribution function in Zel'dovich
approximation as a function of the initial relative velocity
distribution assuming that it is Gaussian.  Then in \S 3 we calculate 
it explicitly for the standard cold-dark-matter (CDM) model
spectrum.  We first trace its time evolution and then compare our
results corresponding to the present epoch with $N-$body simulations
done by ZQSW.  Finally \S 4 is devoted to discussion and 
conclusion.

\section{Pairwise velocity distribution in Zel'dovich approximation}

In the Zel'dovich approximation (ZA), the Eulerian position, $\vex (t)$, of 
a mass element is given by a function of its Lagrangian or initial
coordinate $\veq$ as,
\beq
  \vex(t)=\veq+D(t)\vep(\veq),  \label{xt}
\eeq
where $\vep(\veq)$ is a shift vector and $D(t)$ is a growth factor.
The density contrast, $\delta(\vex,t)$, reads
\beq
   \delta(\vex,t)=\left|\frac{\partial (x_1,x_2,x_3)}{ \partial (q_1,q_2,q_3)}
   \right|^{-1}-1 \simeq -D(t)\nabla_\vesq\cdot\vep(\veq),
\eeq
where the latter approximate equality holds when $\delta$ is small.
$D(t)$ is usually identified with the growth factor of the linear
growing mode, and we do so here, too, but it can be an arbitrary linear 
combination of linear growing and decaying modes in principle.
Then from equation (\ref{xt}) the peculiar coordinate velocity 
also has only the growing
mode in the former case;
\beq
   \vev(t)\equiv \dot{\vex}(t)=\dotD(t)\vep(\veq),
\eeq
so it should be  identified with the growing longitudinal velocity.

Let us consider time evolution of elements $A$ and $B$ with the
Lagrangian coordinates $\veqa$ and $\veqb\equiv\veqa+\ver$, respectively.
The Eulerian position and peculiar velocity at each point read
\beqa
  \vexa(t)\equiv\vex(\veqa,t)&=&\veqa+D(t)\vep(\veqa)\equiv\veqa+D(t)\vepa,
  \nonumber\\
  \vexb(t)\equiv\vex(\veqb,t)&=&\veqb+D(t)\vep(\veqb)\equiv\veqb+D(t)\vepb,
  \\
  \veva(t)\equiv\vev(\veqa,t)&=&\dotD(t)\vepa, \nonumber\\
  \vevb(t)\equiv\vev(\veqb,t)&=&\dotD(t)\vepb. \nonumber  
\eeqa
We investigate time evolution of their relative
velocity,
\beqa
  \vevab(t)&\equiv&\vevb(t)-\veva(t)=\dotD(t)(\vepb-\vepa) \nonumber\\
  &\equiv& \vevp(t)+\vevv(t),
\eeqa
where $\vevp(t)$ and $\vevv(t)$ represent components parallel and
perpendicular to $\verab(t)\equiv\vexb(t)-\vexa(t)$. 
They read
\beqa
  \vevp(t)&=&\frac{\vevab(t)\cdot\verab(t)}{|\verab(t)|^2}\verab(t)
    \equiv\vp(t)\frac{\verab(t)}{\rab(t)},\\
  \vevv(t)&=&\vevab(t)-\vevp(t),\nonumber
\eeqa 
respectively, where
$\rab(t)$ and $\vp(t)$ are  given by
\beqa
  \rab^2(t)&=&\lmk r+\frac{D}{\dotD_i}\vpi\rmk^2 
  +\lmk\frac{D}{\dotD_i}\vvi\rmk^2, \\
  \vp(t)&=&\frac{\dotD}{\rab(t)}\lmk\frac{r\vpi}{\dotD_i}
  +\frac{D\vabi^2}{\dotD_i^2}\rmk.
\eeqa
Here suffix $i$ denotes quantities at the initial time $t=t_i$ when
$D(t_i) \equiv D_i$ was negligibly small.

Thus one can calculate the probability distribution function (PDF),
$P(V;R,t)$, of
pairwise peculiar velocity $\vp(t)=V$ with separation $\rab(t)=R$
from the initial PDF, $p(\vpi,v_{\bot xi},v_{\bot yi};r)$, as
\beq
P(V;R,t)=\frac1{4\pi R^2}\int 4\pi r^2 dr dv_{\| i} dv_{\bot xi}dv_{\bot yi}
p(\vpi,v_{\bot xi},v_{\bot yi};r)\delta(R-\rab(t))\delta(V-\vp(t)).\lab{def}
\eeq
Here $v_{\bot xi}$ and $v_{\bot yi}$ are the two components of
$\vev_{\bot i}$ vertical to each other.
 From the global isotropy the initial PDF depends on $\vpi$ and
$v_{\bot i}\equiv \sqrt{v_{\bot xi}^2+v_{\bot yi}^2}$ only.  
We find after some manipulation,
\beq
P(V;R,t)=\frac{2\pi\dotD_i^3}{D^2\dotD}\int_{r^*}^{\infty}rdr
p(\vpi^*,\vvi^*;r),\label{kiso}
\eeq
with
\beqa
  \vpi^*&\equiv&\frac{\dotD_i}{D}\lkk \frac{R}{r}
  \lmk R - \frac{D}{\dotD}V\rmk -r \rkk, \\
  \vvi^*&\equiv&\frac{\dotD_i}{D}R\lkk 1-\frac{1}{r^2}
  \lmk R-\frac{D}{\dotD}V\rmk^2\rkk^{\frac{1}{2}}, 
\eeqa
and
\beq
  r^* \equiv \left| R-\frac{D}{\dotD}V\right|. \label{rmin}
\eeq

In order to evaluate the desired PDF let us specify the initial PDF, 
$p(\vpi^*,\vvi^*;r)$.  Since we are dealing with only the
longitudinal mode, in the initial or linear regime 
the peculiar velocity is related with
$\delta(\vex,t)$ or its Fourier
transform, $\deltak(t)$, as 
\beq
  \vev(\vex,t_i)=i\frac{\dotD_i}{D_i}\int\frac{\vek}{k^2}\deltak(t_i)
  e^{i\vesk\vesx}\frac{d^3k}{(2\pi)^{\frac{3}{2}}},
\eeq
(Peebles 1980).
Assuming the initial density fluctuation is random Gaussian with
the power spectrum $P_i(k)$ defined by
\beq
     \langle \deltak(t_i)\delta_{\vesk'}^{*}(t_i)\rangle = P_i(k)(2\pi)^3
     \delta(\vek - \vek'),
\eeq
we find the initial
pairwise peculiar velocities are also Gaussian distributed with the
two-point correlation functions (G\'orski 1988),
\beqa
  \llangle \vpi \vpi \rrangle &=& \frac{8\pi}{3}\lmk\frac{\dotD_i}{D_i}\rmk^2
  \int dk P_i(k)\lkk 1-3j_0(kr)+6\frac{j_1(kr)}{kr}\rkk \equiv Q_\|(r), \\
\llangle v_{\bot xi} v_{\bot xi} \rrangle&=&
\llangle v_{\bot yi} v_{\bot yi} \rrangle =
\frac{8\pi}{3}\lmk\frac{\dotD_i}{D_i}\rmk^2
\int dk P_i(k)\lkk 1- 3\frac{j_1(kr)}{kr}\rkk \equiv Q_\bot (r).  \label{bv} 
\eeqa
One can also express $Q_\|(r)$ and $Q_\bot (r)$ in terms of the
integral of the initial two-point correlation function, $\xii(r)$,
as
\beqa
 Q_\|(r)&=&\frac{2}{3}\lmk\frac{\dotD_i}{D_i}\rmk^2
\lkk J_2(r)-\frac{J_5(r)}{r^3}\rkk \equiv
\lmk\frac{\dotD_i}{D_i}\rmk^2
S_\|(r),\\
Q_\bot (r)&=&\lmk\frac{\dotD_i}{D_i}\rmk^2
\lkk \frac{2}{3}J_2(r)-\frac{J_3(r)}{r}+\frac{J_5(r)}{3r^3}\rkk
\equiv \lmk\frac{\dotD_i}{D_i}\rmk^2 S_\bot (r),
\eeqa
 where
\beq
  J_n(r)\equiv \int_0^r\xii(r)r^{n-1}dr,
\eeq
(Peebles 1993, \S 21).
We thus find 
\beq
p(\vpi^*,\vvi^*;r)=\frac{e^{-W}}{\sqrt{(2\pi)^3Q_\|(r)Q_\bot^2(r)}}, ~~~~~
W\equiv\frac{( \vpi^* )^2}{2Q_\|(r)}+\frac{( \vvi^* )^2}{2Q_\bot(r)}.\lab{linp}
\eeq
We can therefore obtain the desired PDF,
$P(V;R,t)$, through the integration in equation (\ref{kiso}) if we specify
$\xii(r)$ or $P_i(k)$.  The final expression reads
\beq
  P(V;R,t)=\frac{1}{\sqrt{2\pi}}\frac{D}{\dotD}\lmk\frac{D_i}{D}\rmk^3 
 \int_{r^*}^{\infty}rdr \frac{e^{-W}}{\sqrt{S_\|(r)S_\bot^2(r)}},
  \label{final}
\eeq
\[
 W =  \frac{1}{2}\lmk\frac{D_i}{D}\rmk^2\lnk \frac{1}{S_\|(r)}
 \lkk \frac{R}{r}\lmk r - \frac{D}{\dotD}V\rmk - r \rkk^2
 + \frac{R^2}{S_\bot(r)}\lkk 1- \frac{1}{r^2}\lmk r -
\frac{D}{\dotD}V\rmk^2 \rkk\rnk.  \nonumber
\]
We readily see that it is independent of $\dotD_i$.  
Therefore we can deal with an extreme case peculiar velocity is set to 
be zero initially as in ZQSW without any difficulty.  
Furthermore,
since $S_\|(r)$ and $S_\bot(r)$ are proportional to $D_i^2$ in the
linear regime when the initial condition is set, equation (\ref{final}) is 
also independent of $D_i$.

Since $S_\|(r)$ and $S_\bot (r)$ approach the same value
\beq
  S_\infty \equiv \frac{8\pi}{3}\int dk P_i(k),
\eeq
in the limit $r \longrightarrow \infty$, we can find an analytic
expression of $P(V;R,t)$ in the case $r^*$ is much larger than the
typical correlation scale as
\beq
  P(V;R,t) \simeq \frac{1}{\sqrt{2\pi S_\infty}}\frac{D_i}{\dotD}
  \exp \lkk -\frac{V^2}{2S_\infty}\lmk\frac{D_i}{\dotD}\rmk^2\rkk
  \equiv P_G(V;t),  \label{Gaussiantail}
\eeq
independent of $R$.  Thus 
the PDF as calculated from ZA (eq.\ [\ref{final}]) approaches Gaussian
(eq.\ [\ref{Gaussiantail}]) not only for a large separation, $R$,  
but also in the
cases $V$ or $D$ is large.  The latter cases, however, correspond to a 
highly nonlinear regime where the Zel'dovich approximation is not reliable.

Note that $p(\vpi^*,\vvi^*;r)$ in equation 
(\ref{linp}) has been normalized so that
velocity integral of $P(V;R,t)$ in equation (\ref{final}) results in the
two-point correlation function of the evolved density field,
$\xi(R,t)$, as
\beq
  \xi(R,t)=\int_{-\infty}^\infty P(V;R,t)dV -1,  \label{corr}
\eeq
(Bharadwaj 1996).

\section{Pairwise velocity distribution in CDM model}
\subsection{CDM power spectrum} \label{cdmspect}
In this section, we calculate the equation (\ref{final}) with a
specific model.
We adopt the standard CDM model as a typical model of structure
formation which has also been employed by ZQSW, namely, 
consider the Einstein de 
Sitter universe with the current Hubble parameter 
$H_0=50{\rm km/s/\Mpc}$.  We also normalize the scale factor 
 $a=1$ at present and set $D=a$. 
The initial power spectrum of density fluctuation is taken from
Efstathiou \etal (1992) as 
\beq
  P_i(k)=\frac{BD_i^{2}k}{\lnk 1 + 
  \lkk \alpha k + (\beta k)^{\frac{3}{2}} + (\gamma k)^2 
  \rkk^\nu\rnk^{\frac{2}{\nu}}},
        \label{CDM}
\eeq
with $\alpha =25.6$ Mpc, $\beta =12$ Mpc, $\gamma =6.8$ Mpc, 
and $\nu=1.13$. $B$ is the
normalization factor which may be determined from the anisotropy of
microwave background radiation as 
\beq
  B=\frac{12 \Omega_0^{-1.54}}{5\pi H_0^4}\lmk\frac{Q_{rms}}{T_0}\rmk^2,
\eeq 
where $Q_{rms}$ is the quadrupole fluctuation amplitude and $T_0=2.73$K is
the current temperature.  The four-year COBE data gives $Q_{rms-PS}\simeq
15.3\mu$K (G\'orski \etal 1996), while it
ranges between $Q_{rms}= 9.5\mu$K and $23.2\mu$K in the simulations of ZQSW.

\subsection{typical time evolution}
First we show how the distribution function of a fixed comoving
separation, $R$, evolves in time. Figure 1(a) depicts evolution of
$P(V;R,t)$ for the comoving scale corresponding to $R=2$ Mpc today 
at $D=0.05,~0.1,~0.2,~0.5,~1.0$ and 10.0.  The
horizontal velocity axis is normalized by the
one point velocity dispersion proportional to
$D\dot{D}$ and the numbers on the velocity axis apply for $D=1$ only.
Figure 1(b) represents time evolution for the comoving scale $R=6$ Mpc 
today.
These two figures imply that 
the negative tail develops much faster than the positive 
part  from symmetric Gaussian
distribution, properly reflecting the attractive nature of gravitational
interaction.  
In the intermediate stage of the evolution around $D=0.5~-~1$ for Fig.\ 
1(a) and $D\sim 1$ for Fig.\ 1(b), an
exponential shape is prominent around the peak.  We thus find that the 
exponential pairwise peculiar velocity distribution can be realized as a
result of semi-nonlinear dynamics of the Lagrangian perturbation theory.

In ZA, however, each mass element moves
kinematically (eq.\ [\ref{xt}]) even after shell crossing, hence structures
or correlations are destroyed as we extrapolate the evolution with it.  As is
seen in Figs.\ 1 the distribution becomes practically 
Gaussian again by the epoch $D=10$ except around $V=0$.  
In this highly nonlinear regime ZA
is by no means valid.  We thus see that non-Gaussian shape of the
distribution function $P(V;R,t)$ with a nearly exponential form near
the peak is realized in this approximation as a transient phenomenon
in the semi-nonlinear regime.

Finally we briefly comment on the evolution of the 
two-point correlation function of the
density field $\xi(R;t)$ obtained from equation (\ref{corr}).
On the comoving  $R=6$ Mpc scale it evolves from $0$ to $0$  with the
maximum value of $1.0$ at $D=1.2$.  ZA predicts smaller amplitude 
of $\xi(R;t)$ than $N-$body simulations  in the nonlinear
regime (Schneider \& Bartelmann 1995, Wei\ss\  \etal 1996).  We return
to this point below. 

\subsection{Dependence on the separation}

Next we set $D=1$ and analyze dependence on the separation $R$.
We plot the new distribution function for the CDM spectrum 
 normalized by COBE data, $Q_{rms}=15.3\mu$K, 
in Fig.\ 2(a) on scales
$R=2,~ 6$, and $10$ Mpc.  In Fig.\ 2(b)
the radial velocity distribution given by the linear 
theory is depicted for comparison.
Even on the distance of $10$ Mpc, these two distributions are
significantly different from each other.

As mentioned in the last subsection, with the original CDM power spectrum ZA 
predicts smaller magnitude of
the correlation function in the nonlinear regime due to the fact that
mass elements continue streaming even after shell crossing without
getting bounded and virialized.  To cure this
problem it has been suggested in the literatures (Coles \etal 1993,
see also Kofman \etal 1992) 
to adopt the
``truncated'' Zel'dovich approximation (TZA), in which 
small-scale power of the initial density fluctuation is
phenomenologically suppressed using an
appropriate window function such as a Gaussian filter and then apply
ZA to the processed power spectrum, 
\beq
   P_t(k)=\exp(-k^2R_f^2)P_i(k),
\eeq
where $R_f$ is the filtering length to be determined so that the
results agree with the predictions of $N-$body simulations. 
Here we choose the filtering scale $R_f=2.2$ Mpc so that the amplitude
of the two-point correlation function at $R=6$ Mpc becomes equal to that
of the simulation (ZQSW). 
Figure 2(c) is the distribution function based on TZA where we find
too small velocity dispersion.  
The filtering length we have adopted here is a modest one.  In fact,
for the purpose of optimizing the overall performance of ZA in
reproducing the clustering properties of  CDM $N-$body simulations,
an even larger value, $R_f=5.5$ Mpc, has been suggested by Schneider
and Bertelmann (1994).  Then the magnitude of velocity dispersion
would be even smaller.
We thus find that TZA mimics
properties of actual matter distribution by suppressing the velocity
field and so it is not appropriate to analyze pairwise velocity field 
with this approximation.

Next we compare our results with the numerical results 
of ZQSW who plotted frequency distributions of pairwise velocity for
the pairs with separations 1 -- 2 Mpc, 5 -- 6 Mpc, and 10 -- 11 Mpc with the
same condition as the figures of ZQSW with $Q_{rms}=9.5\mu$K.
In Fig.\ 3(a) we plot our result for $R=$1 Mpc (dotted line) and 2 Mpc
(solid line) as well as the Gaussian tail distribution function
$P_G(V;t)$ defined in equation (\ref{Gaussiantail}) by short-dashed line. 
Also depicted there is a long-dashed line which mimics the
histogram of ZQSW near the peak where exponential ({\it i.e.} linear
in the figures) fit is good.  This line has been drawn from eye-fitting
of Fig.\ 7(a) of ZQSW.  Similarly in Figs.\ 3(b) and 3(c) results
for $R=$5 and 6 Mpc and $R=10$ and 11 Mpc are depicted together with
approximate curves inferred from the  $N-$body experiment of ZQSW.

As is seen there we observe a systematic deviation to right in all the 
three scales.  This is because gravitational infall is not completely
taken into account in ZA.  In fact the average infall velocities
calculated by ZQSW are $\langle V_{\rm num} \rangle=-280$km/s, $-355$km/s, and
$-276$km/s for the pairs with $R=1$ -- 2 Mpc, 5 -- 6 Mpc, and 10 -- 11 Mpc,
respectively, while ZA gives $\langle V_{\rm ZA} \rangle=14$km/s, 
$-0.49$km/s, 
and $-69$km/s for $R=$2, 6, and 11  Mpc, respectively.  If we correct 
this deviation of average infall by shifting the origin of the
velocity axis with an appropriate amount,
 we find that the distribution function
calculated with ZA reproduces the results of numerical simulation well 
especially near the peak in Figs.\ 3(b) and 3(c).

This means that even if ZA fails to reproduce the average infall, it
predicts pairwise velocity dispersion, $\sigma_V(R)$, fairly close to
the corresponding numerical values.  For an symmetric exponential
distribution, the velocity dispersion appears in the PDF as
\beq
  P(V;R)dV \propto \exp\lkk - \frac{\sqrt{2}|V|}{\sigma_V(R)}\rkk dV. 
  \label{exponential}
\eeq
Since the PDF obtained from $N-$body simulations is asymmetric, it
would be more appropriate to treat left- and right-hand-sides of the
peak separately.  Applying the fitting function of the type of
equation 
(\ref{exponential}) independently on each side of the peak in Figs.\
3(a)--3(c), we estimate the velocity dispersion on each side, 
$\sigma_{V\rm left,right}(R)$, from the slope of each long-dashed line
which is shown in the first column of Table 1 named  Figure.
Also shown there are the results based on ZA.  The second column, ZA2, 
is obtained from the least-square fit between the theoretical curve
(eq.\ [\ref{final}]) and equation (\ref{exponential}) 
in the region $P(V;R) < e^{-2}
\max P(V;R)$ where the two sides are again treated independently.
The third column, ZA1.5, is the same as ZA2 except for the narrower
fitting region $P(V;R) < e^{-1.5}\max P(V;R)$.  In both cases the
central flatter region has been removed from fitting for $R=11$ Mpc
just as we have done so in drawing long-dashed curves in Fig.\ 3(c).

As is seen there, ZA reproduces the asymmetric feature of the slope on
both sides qualitatively well and on larger scale the agreement is
even quantitative.

\section{Discussion}

In the present paper we have formulated the pairwise peculiar velocity 
PDF with ZA starting from Gaussian fluctuations.  Using the standard
CDM power spectrum we have calculated the PDF on various scales at
different epochs.  As a result we have found that even if the
one-point velocity distribution remains Gaussian, non-Gaussian feature 
develops prominently and we find an exponential shape on megaparsec 
scales before and around the epochs corresponding to the present.

One may be tempted to calculate the relative 
velocity PDF of pairs of higher density
regions or density peaks because the numerical data of ZQSW are those
of halos where galaxies are likely formed.  In ZA the density
contrast is a function of $\partial p_i/\partial q_j$ while velocity
field is that of $p_i$, and since they have no correlations in the
isotropic Gaussian distribution, velocity PDF is independent of the
configuration of the density field.

Previous explanations on the exponential PDF (Sheth 1996, Diaferio \&
Geller 1996) are based on highly nonlinear dynamics of galactic
systems.  That is, the exponential feature appears as a result of
superposition of Gaussian velocity distributions in clumps with
various velocity dispersions.  Hence although this model
 is suitable to account for a symmetric exponential 
distribution with vanishing mean net relative velocity,
a more elaborate treatment is necessary in order to reproduce an asymmetric
exponential distribution with its peak at a negative relative
velocity as observed in numerical simulations.

The PDF based on ZA naturally realizes an asymmetric exponential
feature, although it also failed to predict the location of the peak
correctly at $D=1$.
In fact, as discussed in the literatures (Coles \etal 1993, Schneider
\& Bertelman 1994, Wei\ss\ \etal 1996) ZA does not successfully
reproduce clustering properties of density field in the regime $D
\simeq 1$, and it has been suggested to use the truncated spectrum to
make the agreement better for $D \gsim 0.2$ (Schneider \& Bertelman 1994).
On the other hand, as we have shown in Fig.\ 2(c), this TZA predicts
too small amplitudes of the velocity dispersion.  Thus neither the
original ZA nor TZA reproduces both density distribution and pairwise
velocity distribution simultaneously for the present universe.  Hence
it is not very certain to which extent the semi-nonlinear dynamics
discussed here plays a role in realizing the actual exponential
distribution in the present universe.  In order to draw a more
quantitative conclusion, it is desirable to compare our formula
(eq.[\ref{final}]) with the time evolution of $N-$ body simulations of a CDM 
model, especially around the epoch $D \simeq 0.2$ up to when the
original ZA suffices for reproducing the power spectrum of density
field.  Unfortunately, we have been unable to do so due to the lack of 
published data with the sufficient informations about the initial
condition, the epoch, and the separation in physical units.

Our conclusion is therefore limited to a qualitative one.  Nonetheless 
we believe it worth pointing out that simple kinematics as ZA results
in an exponential shape of the pairwise peculiar velocity
distribution.  Indeed the pairwise PDF, which contains much more
information on clustering in that its integral gives the two-point
correlation function, is totally different from the
one-point PDF.

\acknowledgments

We are grateful to Professor W.H.\ Zurek for a correspondence.
NS would like to thank Professor
N. \ Sugiyama for useful comments and  Professor H.\ Sato
for his continuous encouragements. He also
acknowledges support from JSPS.
This work was partially supported by the Japanese Grant
in Aid for Science Research Fund of the Ministry of Education, Science,
Sports and Culture  No.\ 3161(NS) and No.\ 09740334(JY).

\newpage
\centerline{\bf FIGURE CAPTIONS}

\smallskip

\begin{description}
  
\item[Figs.\ 1] Evolution of the radial pairwise velocity distribution
    $P(V;R,t)$ on the comoving separations $R=2$ Mpc (a) and $R=6$ Mpc 
(b) in terms of ZA
with the CDM initial power spectrum with $Q_{rms}=15.3\mu {\rm K}$. 
Solid line represents the distribution at $D=0.05$, dotted
line at $D=0.1$,  
short-dashed line at $D=0.2$, long-dashed line at $D=0.5$,  dash-dotted
line at $D=1$ (corresponding to present epoch), and thick solid line at
$D=10.0$.
To compare the statistical shape at different epochs, 
the velocity axis has been normalized by the
one-point velocity dispersion proportional to
$D\dot{D}$ and the numbers on the velocity axis apply for $D=1$ only.
  
\item[Figs.\ 2]The distribution of radial pairwise distribution $P(R;V,t)$
    on the separations $R=2$ Mpc (thin solid line), $R=6$ Mpc 
(dotted line), $R=10$ Mpc (dashed line) and $R=+\infty$ (thick solid 
line) at present, $D=1$.
(a) represents the results based on ZA with the full CDM spectrum with 
$Q_{rms}=15.3\mu$K,
(b) is based on the linear theory with the same spectrum,
(c) is obtained from ZA but with the truncated CDM power spectrum with the
filtering length $R_f=2.2$ Mpc.
  
\item[Figs.\ 3]Schematic comparison of our formula with the results of 
    $N-$body simulation of 
    ZQSW for three typical separations at $D=1$ with $Q_{rms}=9.5\mu$K. 
    In each figure, solid and dotted lines are the PDF obtained from ZA.
    The long-dashed line mimics the result of ZQSW in the regions an
    exponential fit is reasonable.
    (a) is for the pairs of the separation $R=1~-~ 2{\rm Mpc}$, 
    (b) for $R=5~-~6{\rm Mpc}$ and (c) for $R=10~-~ 11{\rm Mpc}$.
     The short-dashed line is the Gaussian distribution $P_G(V;t)$ 
    defined in equation (\ref{Gaussiantail}).
\end{description}
\newpage
\begin{center}
TABLE 1\\
{\sc Velocity Dispersion Obtained from the Slope of the Exponential
Fitting Function}\\
\ \\
\begin{tabular}{ccccc}
\hline\hline
 $R$ (Mpc) & Side  & Figure$^a$ ($10^2$km/sec) & ZA2$^b$ ($10^2$km/sec) &
ZA1.5$^c$ ($10^2$km/sec) \\
 2 ~~ &~~ left~~ &~~ 8.3~~ &~~ 5.7~~ &~~ 5.9~~ \\
 2 ~~ &~~ right~~ &~~ 4.3~~ &~~ 4.5~~ &~~ 4.5~~ \\
 6 ~~ &~~ left~~ &~~ 8.5~~ &~~ 7.0~~ &~~ 8.1~~ \\
 6 ~~ &~~ right~~ &~~ 4.2~~ &~~ 5.1~~ &~~ 5.5~~ \\
11~~ &~~ left~~ &~~ 7.0~~ &~~ 7.0~~ &~~ 8.2~~ \\
11~~ &~~ right~~ &~~ 3.4~~ &~~ 4.1~~ &~~ 3.5~~ \\
\hline
\end{tabular}
\end{center}
$^a$Based on the long-dashed lines in Figs.\ 3(a)--(c).
$^b$From the least-square fit between the theoretical curve
(eq.\ [\ref{final}]) and equation (\ref{exponential}) 
in the region $P(V;R) < e^{-2}\max P(V;R)$. 
$^c$Sane as ZA2 but with a narrower fitting region, 
$P(V;R) < e^{-1.5}\max P(V;R)$.

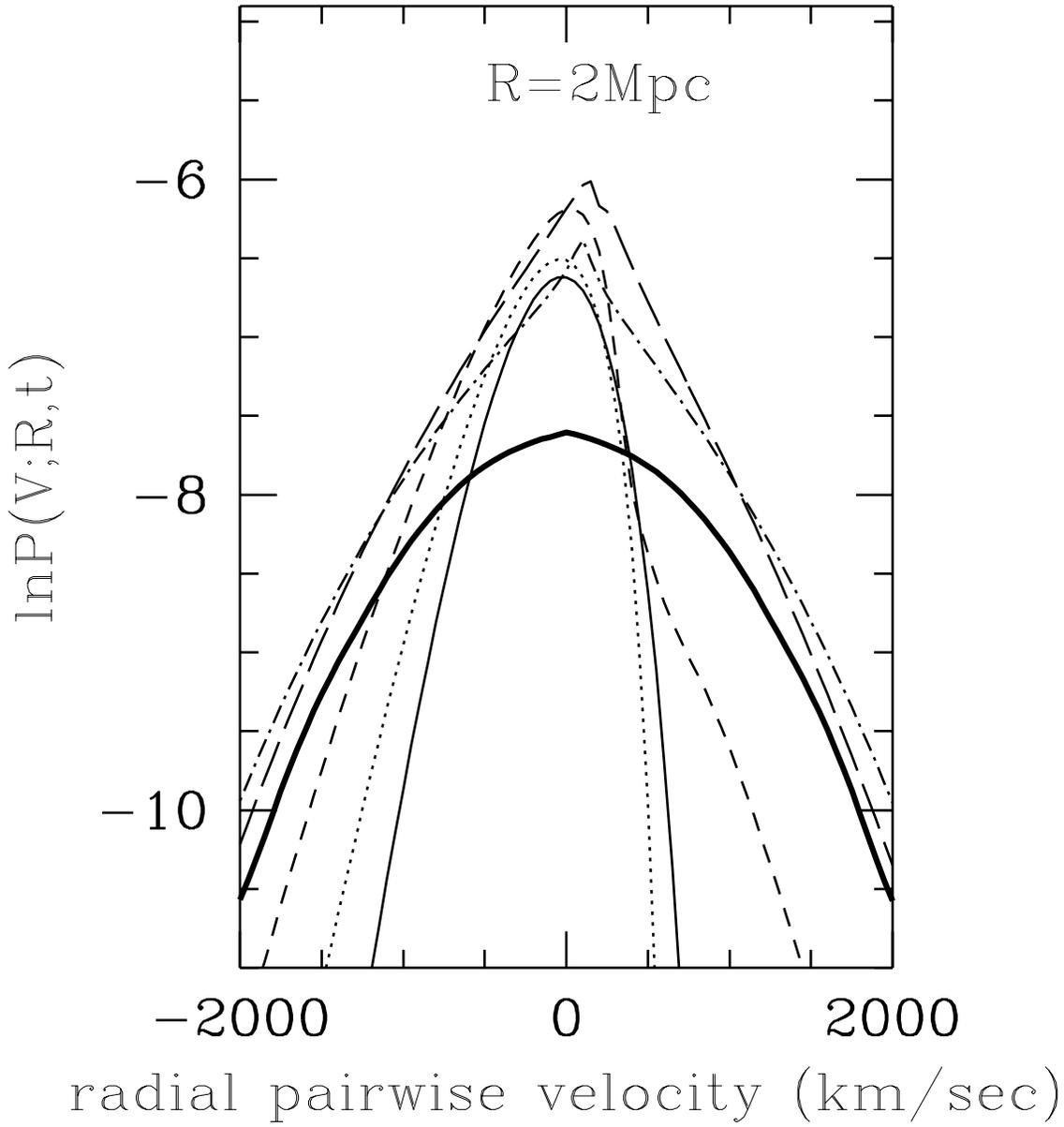
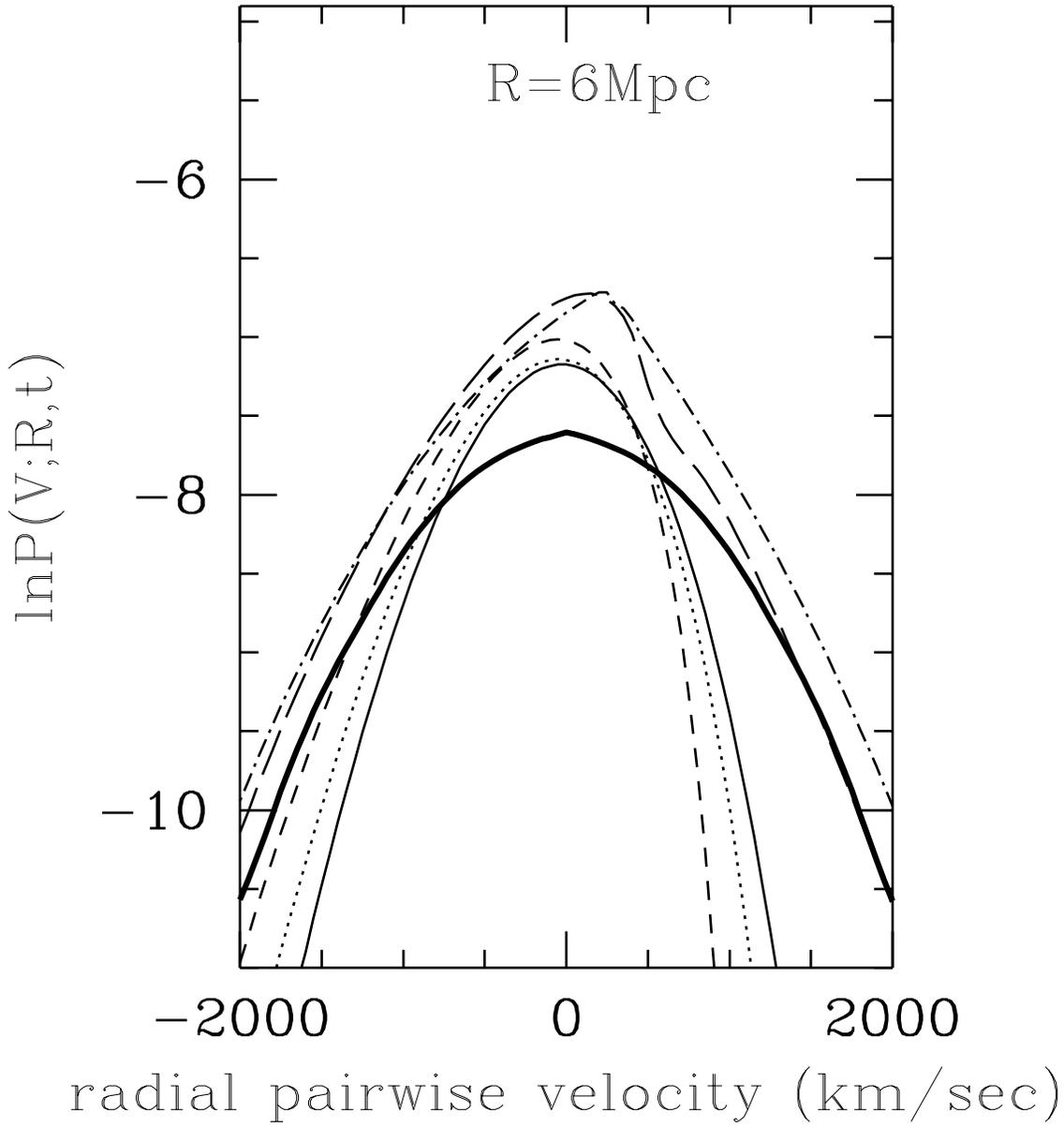
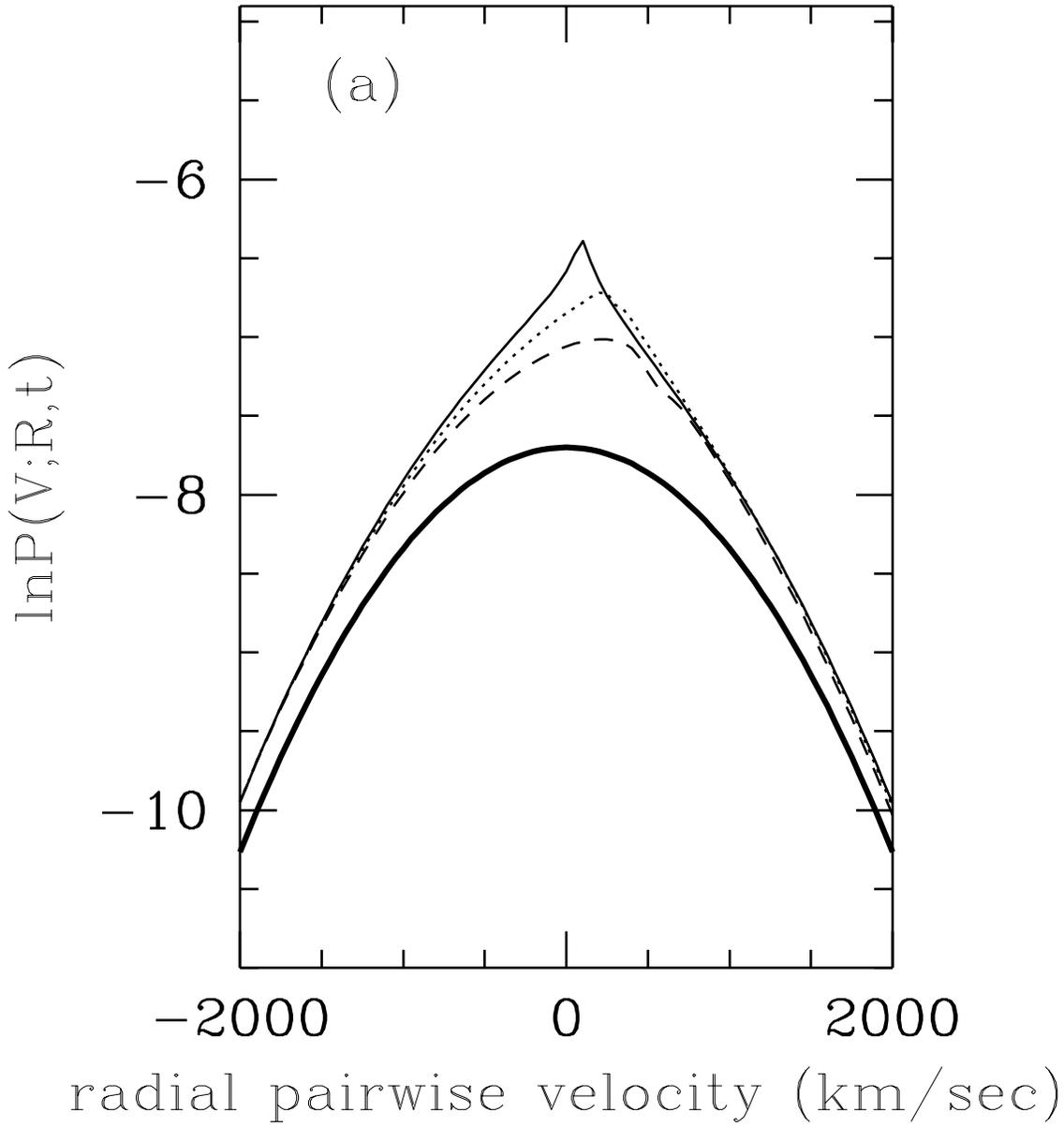
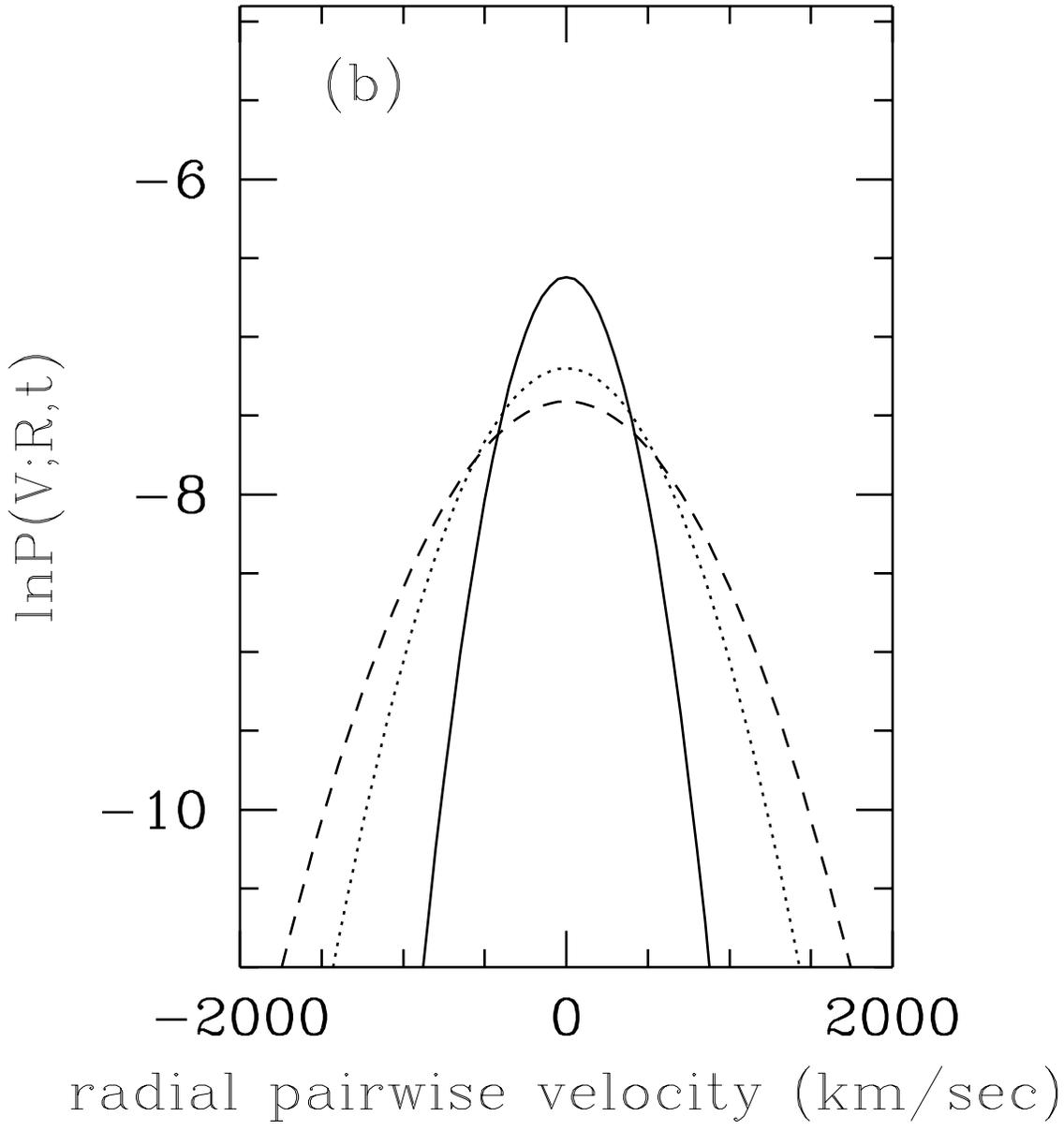
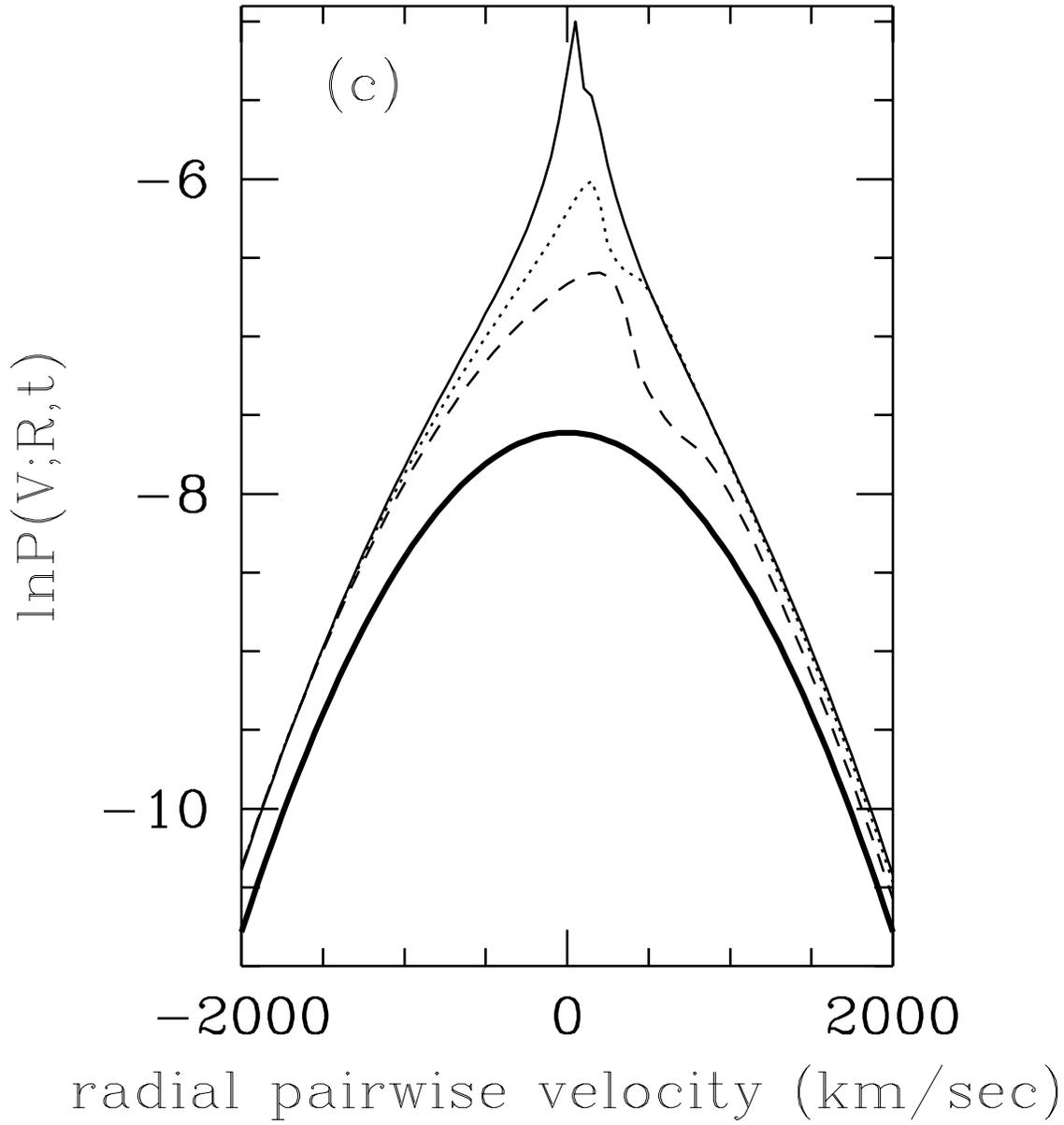
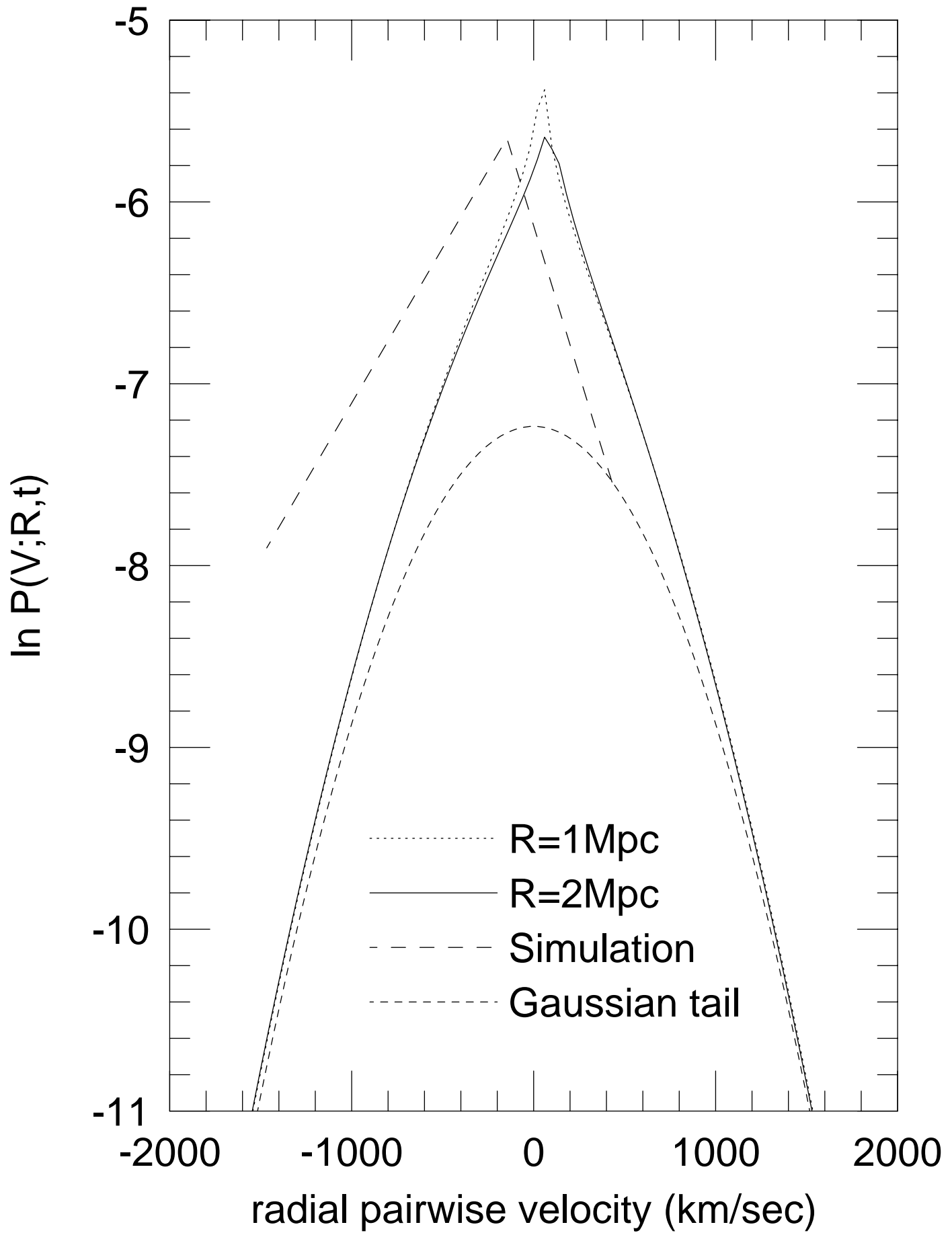
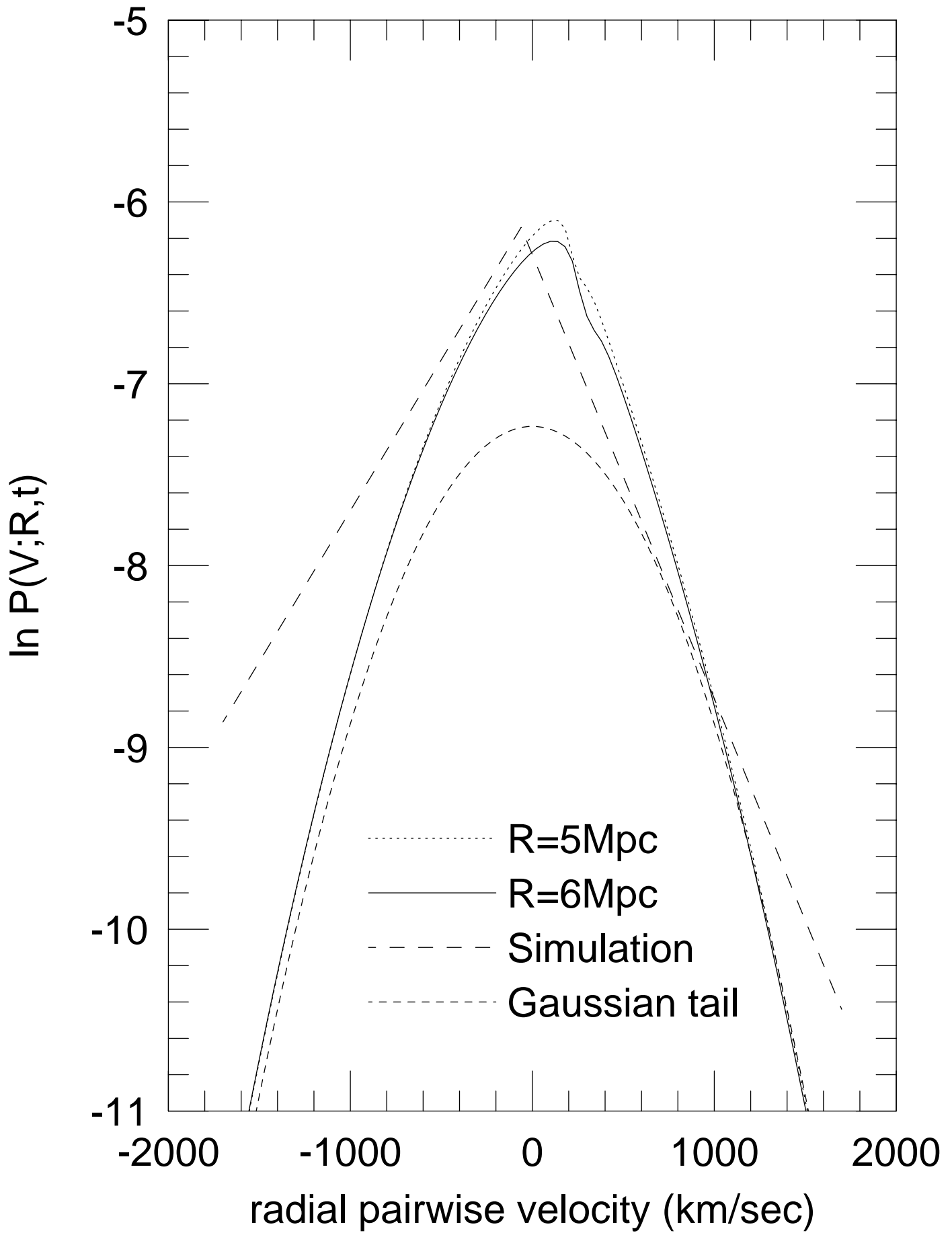
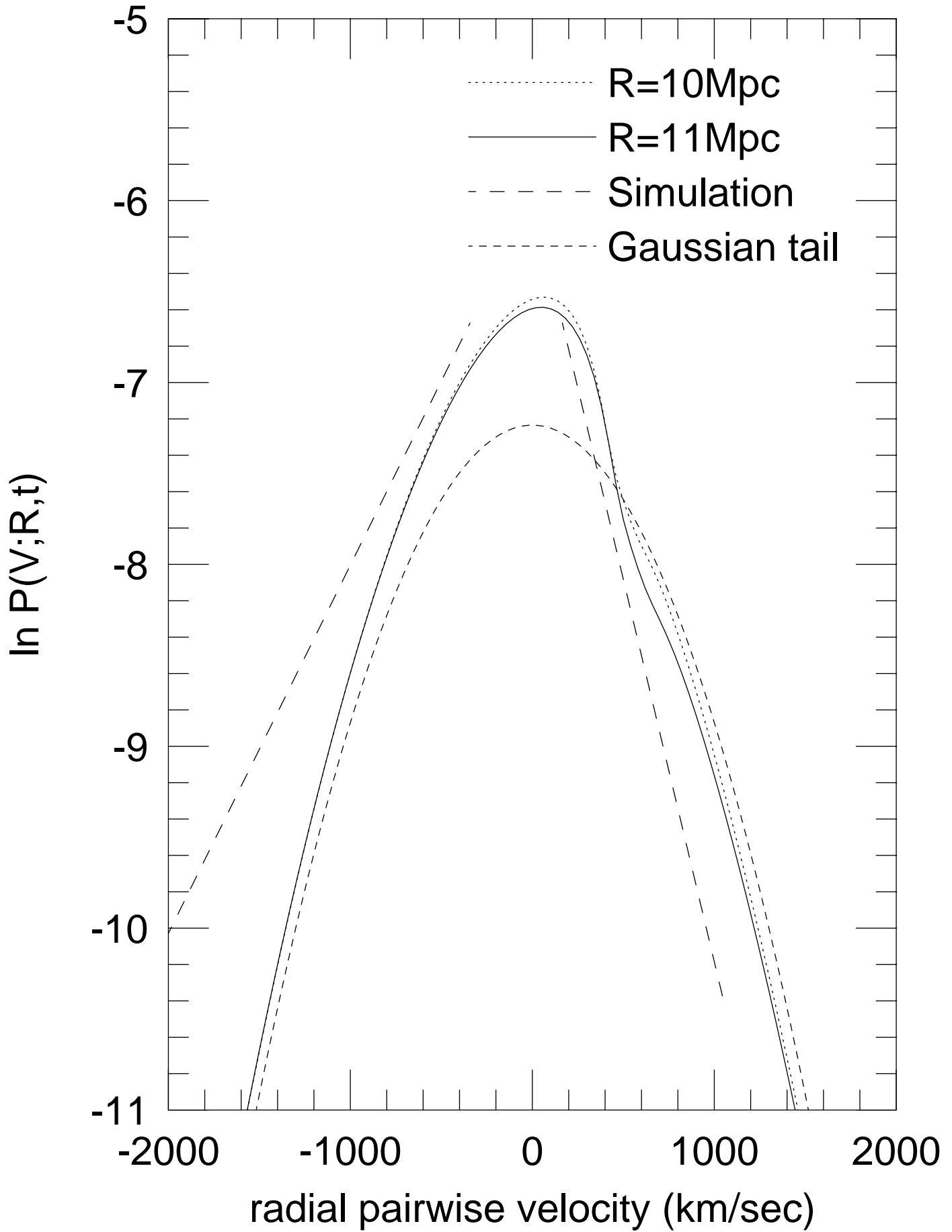
\end{document}